\documentclass{tlp}
\submitted {5th October 2009}
\revised{1th March 2010}
\accepted{21st February 2011}

\input epsf

\newcommand{\ignore}[1]{}

\begin{document}

\long\def\comment#1{}

\title{The Language Features and Architecture of B-Prolog}

\author[N.F. Zhou]
{Neng-Fa Zhou \\
Department of Computer and Information Science \\
CUNY Brooklyn College \& Graduate Center \\
zhou@sci.brooklyn.cuny.edu 
}

\pagerange{\pageref{firstpage}--\pageref{lastpage}}
\volume{\textbf{10} (3):}
\jdate{March 2002}
\setcounter{page}{1}
\pubyear{2002}

\maketitle

\label{firstpage}

\begin{abstract}
B-Prolog is a high-performance implementation of the standard Prolog language with several extensions including matching clauses, action rules for event handling, finite-domain constraint solving, arrays and hash tables, declarative loop constructs, and tabling. The B-Prolog system is based on the TOAM architecture which differs from the WAM mainly in that (1) arguments are passed old-fashionedly through the stack, (2) only one frame is used for each predicate call, and (3) instructions are provided for encoding matching trees. The most recent architecture, called TOAM Jr., departs further from the WAM in that it employs no registers for arguments or temporary variables, and provides variable-size instructions for encoding predicate calls. This paper gives an overview of the language features and a detailed description of the TOAM Jr. architecture, including architectural support for action rules and tabling.
\end{abstract}


\section{Introduction}
Prior to the first release of B-Prolog in 1994, several prototypes had been developed that incorporated results from various experiments. The very first prototype was based on the Warren Abstract Machine (WAM) \cite{Warren83} as implemented in SB-Prolog \cite{SBProlog}. In the original WAM, the decision on which clauses to apply to a call is made solely on the basis of the type and sometimes the main functor of the first argument of the call. This may result in unnecessary creation of choice points and repeated execution of common unification operations among clauses in the predicate. The first experiment, inspired by the Rete algorithm used in production rule systems \cite{rete}, aimed at improving the indexing scheme of the WAM. The results from that experiment included an intermediate language named {\it matching clauses} and a Prolog machine named {\it TOAM} (Tree-Oriented Abstract Machine) which provided instructions for encoding tries called matching trees \cite{ZhouTU90}. Several other proposals had been made with the same objective \cite{RoyDW87,HickeyM89,KligerS90}, but these proposed schemes had the drawback of possibly generating code of exponential size for certain programs.

The WAM was originally designed for both software and hardware implementations. In the WAM, arguments are passed through argument registers so that hardware registers can be exploited in native compilers and hardware implementations. In an emulator-based implementation, however, passing arguments through registers loses its advantage since registers are normally simulated. The second experiment, which took place during 1991-1994, was to have arguments passed old-fashionedly through the stack as in DEC-10 Prolog \cite{WAR77}. The result from that experiment was NTOAM \cite{Zhou94}. In this machine, only one frame is used for each predicate call which stores a different set of information depending on the type of the predicate. This architecture was later refined and renamed to ATOAM \cite{Zhou96}.

During the past fifteen years since its first release, B-Prolog has undergone several major extensions and refinements. The first extension was to introduce a new type of frame, called a {\it suspension frame} for delayed calls \cite{Zhou961}. In WAM-based systems, delayed calls are normally stored as terms on the heap \cite{Carlsson87}. The advantage of storing delayed calls on the stack rather than on the heap is that contest switching is light. It is unnecessary to allocate a frame when a delayed call wakes up and deallocate it when the delayed call suspends again. This advantage is especially important for programs where calls wake up and suspend frequently, such as constraint propagators \cite{Zhou06ar}.

A delay construct like freeze is too weak for implementing constraint solvers. New constructs, first delay clauses \cite{Meier93,Zhou98} and then action rules \cite{Zhou06ar}, were introduced into B-Prolog. While these new constructs give significantly more modeling power, they required only minor changes to the architecture: for action rules, one extra slot was added into a suspension frame for holding events.

The action rule language serves well as a powerful and yet efficient intermediate language for compiling constraints over finite-domain variables. A constraint is compiled into propagators defined in action rules that maintain some sort of consistency for the constraint. The availability of fine-grained domain events facilitates programming {AC-4} like propagation algorithms \cite{Zhou06tr}. As propagators are stored on the stack as suspension frames, allocation of frames is not needed to activate propagators and hence context switching among propagators becomes faster.

Another major extension was tabling. Unlike OLDT \cite{Tamaki86} and SLG \cite{Chen96} which rely on suspension and resumption of subgoals to compute fixed points, the tabling mechanism, called {\it linear tabling} \cite{Zhou01,Zhou08tab}, implemented in B-Prolog relies on iterative computation of top-most looping subgoals to compute fixed points.  Linear tabling is simpler, easier to implement, and more space-efficient than SLG, but a naive implementation may not be as fast due to the necessity of re-computation. Optimization techniques have been developed to make linear tabling competitive with SLG in time efficiency by significantly reducing the cost of re-computation \cite{Zhou08tab}. For tabled predicates, a new type of frame was introduced into the architecture.  Recently, the tabling system has been modified to support table modes, which facilitate describing dynamic programming problems \cite{GuoG08,Zhou10tai}.  The PRISM system \cite{Sato09} has been the main driving force for the design and implementation of the tabling system in B-Prolog.

In 2007, B-Prolog's abstract machine was replaced by a new one named TOAM Jr. \cite{Zhou07toamjr}. This switch improved the speed of B-Prolog by over 60\% on the Aquarius benchmarks \cite{ROY90}. The old machine ATOAM, like the WAM, has a very fine-grained instruction set in the sense that roughly each symbol in the source program is mapped to one instruction. This fine granularity is a big obstacle to fast interpretation due to the high dispatching cost commonly seen in abstract machine emulators. The new machine TOAM Jr uses no temporary registers at all and provides variable-size specialized instructions for encoding predicate calls. In WAM-based systems, similar efforts have also been made to specialize and merge instructions to reduce the cost of interpretation \cite{Costa99,bart:wamvariations,NassenCS01,Morales05}.

The memory manager of B-Prolog has also been improved recently. B-Prolog employs an incremental copying garbage collector \cite{Zhou00gc} based on the one proposed for the WAM by Older and Rummell \cite{OlderR92}. Because of the existence of suspension frames on the stack, the garbage collector also reclaims space taken by unreachable stack frames. The memory manager automatically expands the stacks and data areas before they overflow, so applications can run with any initial setting for the spaces as long as the overall demand for memory can be met.

This paper overviews in Section 2 the language features of B-Prolog, gives in Section 3 a detailed description of TOAM Jr., the architecture B-Prolog has evolved into after nearly two decades, and summarizes in Section 4 the changes made to the memory architecture for supporting action rules and tabling.
The reader is referred to \cite{Zhou06ar} for a detailed description of architectural support for action rules and to \cite{Zhou08tab} for a detailed description of the extension of the architecture for tabling.

\section{An Overview of Language Features of B-Prolog}
In addition to the standard Prolog language, B-Prolog offers several useful new features. In this section, we give an overview of them.

\subsection{Matching clauses}
A matching clause is a form of a clause where the determinacy and input/output unifications are denoted explicitly. The compiler translates matching clauses into matching trees and generates indices for all input arguments. The compilation of matching clauses is much simpler than that of normal Prolog clauses because no complex program analysis or specialization or dynamic indexing \cite{CostaSL07} is necessary; also the generated code tends to be more compact and faster. The B-Prolog compiler and most of the library predicates are written in matching clauses. 

A {\it determinate} matching clause takes the following form:
\begin{tabbing}
aa \= aaa \= aaa \= aaa \= aaa \= aaa \= aaa \kill
\> $H, G$ {\tt =>} $B$
\end{tabbing}
where $H$ is an atomic formula, $G$ and $B$ are two sequences of atomic formulas. $H$ is called the head, $G$ the guard, and $B$ the body of the clause. No call in $G$ can bind variables in $H$ and all calls in $G$ must be in-line tests. In other words, the guard must be {\it flat}. For a call $C$, matching rather than unification is used to select a matching clause in its predicate. When applying the matching clause to $C$, the system rewrites $C$ {\it determinately} into $B$. In other words, call $C$ fails once call $B$ fails.

A {\it nondeterminate} matching clause takes the following form:
\begin{tabbing}
aa \= aaa \= aaa \= aaa \= aaa \= aaa \= aaa \kill
\> $H, G$ {\tt ?=>} $B$
\end{tabbing}
It differs from the determinate matching clause '$H, G$ {\tt =>} $B$' in that the rewriting from $H$ into $B$ is nondeterminate, i.e., the alternative clauses will be tried when $B$ fails. In a predicate definition, determinate and nondeterminate matching clauses can coexist.

The following gives an example predicate in matching clauses that merges two sorted lists:
\begin{verbatim}
   merge([],Ys,Zs) => Zs=Ys.
   merge(Xs,[],Zs) => Zs=Xs.
   merge([X|Xs],[Y|Ys],Zs),X<Y => Zs=[X|ZsT],merge(Xs,[Y|Ys],ZsT).
   merge(Xs,[Y|Ys],Zs) => Zs=[Y|ZsT],merge(Xs,Ys,ZsT).
\end{verbatim}
The cons \verb+[Y|Ys]+ occurs in both the head and the body of the third clause. To avoid reconstructing the term, we can rewrite the clause into the following:
\begin{verbatim}
   merge([X|Xs],Ys,Zs),Ys=[Y|_],X<Y => Zs=[X|ZsT],merge(Xs,Ys,ZsT).
\end{verbatim}
The call \verb+Ys=[Y|_]+ in the guard matches {\tt Ys} against the pattern \verb+[Y|_]+.

\subsection{Action rules}
The lack of a facility for programming ``active'' sub-goals that can be reactive to the environment has been considered one of the weaknesses of logic programming. To overcome this, B-Prolog provides action rules for programming agents \cite{Zhou06ar}. An agent is a subgoal that can be delayed and can later be activated by events. Each time an agent is activated, some action may be executed. Agents are a more general notion than delay constructs in early Prolog systems and processes in concurrent logic programming languages in the sense that agents can be responsive to various kinds of events including instantiation, domain, time, and user-defined events. 

An action rule takes the following form:
\begin{tabbing}
aa \= aaa \= aaa \= aaa \= aaa \= aaa \= aaa \kill
\> $H, G$,{\tt \{}$E${\tt \}} {\tt =>} $B$
\end{tabbing}
where $H$ is a pattern for agents, $G$ is a sequence of conditions on the agents, $E$ is a set of patterns for events that can activate the agents, and $B$ is a sequence of arbitrary Prolog goals (called {\it actions}) executed by the agents when they are activated. The sequence of actions $B$ can succeed only once and hence can never leave choice points behind. The compiler replaces $B$ with {\tt once($B$)} if it predicts that $B$ may create choice points. When the event pattern $E$ together with the enclosing braces is missing, an action rule degenerates into a determinate matching clause.

A set of built-in events is provided for programming constraint propagators and interactive graphical user interfaces. For example, {\tt ins($X$)} is an event that is posted when the variable $X$ is instantiated. A user program can create and post its own events and define agents to handle them. A user-defined event takes the form of {\tt event($X,O$)} where $X$ is a variable, called a {\it suspension variable}, that connects the event with its handling agents, and $O$ is a Prolog term that contains the information to be transmitted to the agents. The built-in {\tt post($E$)} posts the event $E$. In the next subsection, we show the events provided for programming constraint propagators.

Consider the following examples:
\begin{verbatim}
    echo(X),{event(X,Mes)}=>writeln(Mes).
    ping(T),{time(T)} => writeln(ping).
\end{verbatim}
The agent {\tt echo(X)} echoes whatever message it receives. For example,
\begin{verbatim}
    ?-echo(X),post(event(X,hello)),post(event(X,world)).
\end{verbatim}
outputs the message {\tt hello} followed by {\tt world}. The agent {\tt ping(T)} responds to time events from the timer {\tt T}. Each time it receives a time event, it prints the message {\tt ping}. For example,
\begin{verbatim}
    ?-timer(T,1000),ping(T),repeat,fail.
\end{verbatim}
creates a timer that posts a time event every second and creates an agent {\tt ping(T)} to respond to the events. The repeat-fail loop makes the agent perpetual.

The action rule language has been found useful for programming coroutining such as {\tt freeze}, implementing constraint propagators \cite{Zhou06ar}, and developing interactive graphical user interfaces \cite{Zhou03:cglib}. Action rules have been used by \cite{Schrijvers06} as an intermediate language for compiling Constraint Handling Rules and by \cite{Zhou11LPNMR} to compile Answer Set Programs.

\subsection{CLP(FD)}
Like many Prolog-based finite-domain constraint solvers, B-Prolog's finite-domain solver was heavily influenced by the CHIP system \cite{HEN89}. The first fully-fledged solver was released with B-Prolog version 2.1 in March 1997. That solver was implemented with delay clauses \cite{Zhou98}. During the past decade, the action rule language has been extended to support a rich class of domain events ({\tt ins($X$)}, {\tt bound($X$)},{\tt dom($X,E$)}, and {\tt dom\_any($X,E$)}) for programming constraint propagators \cite{Zhou06tr} and the system has been enriched with new domains (Boolean, trees, and finite sets), global constraints, and specialized fast constraint propagators. Recently, the two built-ins {\tt in/2} and {\tt notin/2} have been extended to allow positive and negative table (also called extensional) constraints \cite{Zhou09}.

The following program solves the SEND + MORE = MONEY puzzle. The call {\tt Vars in 0..9} is a domain constraint, which narrows the domain of each of the variables in {\tt Vars} down to the set of integers from 0 through 9. The call {\tt alldifferent(Vars)} is a global constraint, which ensures that variables in the list {\tt Vars} are pairwise different. The operator \verb+#\=+ is for inequality constraints and \verb+#=+ is for equality constraints. The call {\tt labeling(Vars)} labels the variables in the given order with values that satisfy all the constraints. Another well-used built-in, named {\tt labeling\_ff}, labels the variables using the so called {\it first-fail principle} \cite{HEN89}.

\begin{verbatim}
   sendmory(Vars):-
      Vars=[S,E,N,D,M,O,R,Y], 
      Vars in 0..9,
      alldifferent(Vars),     
      S #\= 0,
      M #\= 0,
      1000*S+100*E+10*N+D+1000*M+100*O+10*R+E #= 
      10000*M+1000*O+100*N+10*E+Y,
      labeling(Vars).         
\end{verbatim}

All constraints are compiled into propagators defined in action rules. For example, the following predicate defines a propagator for maintaining arc consistency for the constraint \verb-X+Y #= C-. 
\begin{verbatim}
   x_in_c_y_ac(X,Y,C),var(X),var(Y),
      {dom(Y,Ey)}
       =>         
      Ex is C-Ey,        
      domain_set_false(X,Ex).
   x_in_c_y_ac(X,Y,C) => true.
\end{verbatim}
Whenever an inner element {\tt Ey} is excluded from the domain of {\tt Y}, this propagator is triggered to exclude {\tt Ex}, the counterpart of {\tt Ey}, from the domain of {\tt X}. For the constraint \verb-X+Y #= C-, we need to generate two propagators, namely, {\tt x\_in\_c\_y\_ac(X,Y,C)} and {\tt x\_in\_c\_y\_ac(Y,X,C)}, to maintain the arc consistency. Note that in addition to these two propagators, we also need to generate propagators for maintaining interval consistency since {\tt dom(Y,Ey)} only captures exclusions of inner elements, not bounds.
\ignore{event is posted if the excluded value happens to be a bound. Note also that we need to preprocess the constraint to make it arc consistent before the propagators are generated.}

The {\tt dom\_any($X,E$)} event, which captures the excluded value $E$ from the domain $X$, facilitates implementing AC4-like algorithms \cite{mohrhenderson:ai:1986}. Consider, for example, the channeling constraint {\tt assignment($Xs,Ys$)}, where $Xs$ is a list of variables [$X_1,\ldots,X_n$] (called {\it primal}) and $Ys$ is another list of variables [$Y_1,\ldots,Y_n$] (called {\it dual}), and the domain of each $X_i$ and each $Y_i$ ($i=1,...,n$) is $1..n$. The constraint is true iff $\forall_{i,j}(X_i = j \leftrightarrow Y_j = i)$ or equivalently $\forall_{i,j}(X_i \neq j \leftrightarrow Y_j \neq i)$. Clearly a straightforward encoding of the channeling constraint requires $n^2$ Boolean constraints. With the {\tt dom\_any} event, however, we can use only $2\times n$ propagators to
implement the channeling constraint. Let {\tt DualVarVector} be a vector
created from the list of dual variables. For each primal variable {\tt Xi}
(with the index {\tt I}), a propagator defined below is created to handle
exclusions of values from the domain of {\tt Xi}.
\begin{verbatim}
   primal_dual(Xi,I,DualVarVector),var(Xi),
      {dom_any(Xi,J)} 
      =>
      arg(J,DualVarVector,Yj),
      domain_set_false(Yj,I).
   primal_dual(Xi,I,DualVarVector) => true.
\end{verbatim}
Each time a value {\tt J} is excluded from the domain of {\tt Xi}, assume {\tt Yj} is the {\tt J}th variable in {\tt DualVarVector}, then {\tt I} must be excluded from the domain of {\tt Yj}. We need to exchange primal and dual variables and create a propagator for each dual variable as well. Therefore, in total $2\times n$ propagators are needed.

Thanks to the employment of action rules as the implementation language, the constraint solving part of B-Prolog is relatively small (3800 lines of Prolog code and 4500 lines of C code, including comments and empty lines) but its performance is very competitive with other CLP(FD) systems \cite{Zhou06ar}. Moreover, the action rule language is available to the programmer for implementing problem-specific propagators.

\subsection{Arrays and the array subscript notation}
A structure can be used as a one-dimensional array,\footnote{In B-Prolog, the maximum arity of a structure is 65535.} and a multi-dimensional array can be represented as a structure of structures. To facilitate creating arrays, B-Prolog provides a built-in, called  {\tt new\_array($X,Dims$)}, where $X$ must be an uninstantiated variable and $Dims$ a list of positive integers that specifies the dimensions of the array. For example, the call {\tt new\_array(X,[10,20])} binds {\tt X} to a two dimensional array whose first dimension has 10 elements and second dimension has 20 elements. All the array elements are initialized to be free variables.

The built-in predicate {\tt arg/3} can be used to access array elements, but it requires a temporary variable to store the result, and a chain of calls to access an element of a multi-dimensional array. To facilitate accessing array elements, B-Prolog supports the array subscript notation {\tt $X$[$I_1$,...,$I_n$]}, where $X$ is a structure and each $I_i$ is an integer expression. This common notation for accessing arrays is, however, not part of the standard Prolog syntax. To accommodate this notation, the parser is modified to insert a token $^\wedge$ between a variable token and {\tt [}. So, the notation {\tt $X$[$I_1$,...,$I_n$]} is just a shorthand for {\tt $X$$^\wedge$[$I_1$,...,$I_n$]}. This notation is interpreted as an array access when it occurs in an arithmetic expression, an arithmetic constraint, or as an argument of a call to {\tt @=/2}.\footnote{{\tt $X$ @= $Y$} is the same as {\tt $X$ = $Y$} except that when an argument is an array access or a list comprehension (described later), it is evaluated before the unification.}

 In any other context, it is treated as the term itself. The array subscript notation can also be used to access elements of lists. For example, the {\tt nth/3} predicate can be defined as follows:
\begin{verbatim}
    nth(I,L,E) :- E @= L[I].
\end{verbatim}
Note that, for the array access notation {\tt A[I]}, while it takes constant time to access the {\tt I}th element if {\tt A} is a structure, it takes O({\tt I}) time when {\tt A} is a list.

\subsection{Loops with foreach and list comprehension}
Prolog relies on recursion to describe loops. The lack of powerful loop constructs has arguably made Prolog less acceptable to beginners and less productive to experienced programmers because it is often tedious to define small auxiliary recursive predicates for loops. The emergence of constraint programming languages such as CLP(FD) has further revealed this weakness of Prolog as a modeling language. Inspired by ECL$^i$PS$^e$ \cite{Schimpf02} and functional languages, B-Prolog provides a construct, called {\tt foreach}, for iterating over collections, and the list comprehension notation for constructing lists. 

The {\tt foreach} call has a very simple syntax and semantics. For example, 
\begin{verbatim}
    foreach(A in [a,b], I in 1..2, write((A,I))
\end{verbatim}
outputs four tuples {\tt (a,1)}, {\tt (a,2)}, {\tt (b,1)}, and {\tt (b,2)}. The base {\tt foreach} call has the form:
\begin{tabbing}
aa \= aaa \= aaa \= aaa \= aaa \= aaa \= aaa \kill
\> \> {\tt foreach($E_1$ in $D_1$, $\ldots$, $E_n$ in $D_n$, $LocalVars$,$Goal$)} 
\end{tabbing}
\noindent
where {\tt $E_1$ in $D_1$} is called an {\it iterator} ($E_1$ is called the {\it pattern} and $D_i$ the {\it collection} of the iterator), $Goal$ is a callable term, and $LocalVars$ (optional) specifies a list of variables in $Goal$ that are local to each iteration. The pattern of an iterator is normally a variable but it can be any term; the collection of an iterator is a list of terms and the notation {\tt $B_1$..$Step$..$B_2$} denotes the list of numbers $B_1$, $B_1$+$Step$, $B_1$+$2*Step$, $\ldots$, $B_1$+$k*Step$ where $B_1$+$k*Step$ is the last element that does not cross over $B_2$. The notation {\tt $L$..$U$} is a shorthand for {\tt $L$..1..$U$}. The {\tt foreach} call means that for each permutation of values $E_1 \in D_1$, $\ldots$, $E_n \in D_n$, the instance $Goal$ is executed after local variables are renamed. 

In general, a {\tt foreach} call may also take as an argument a list of accumulators that can be used to accumulate values from each iteration. With accumulators, we can use {\tt foreach} to describe recurrences for computing aggregates. Recurrences have to be read procedurally.  For this reason, we adopt the list comprehension notation for constructing lists declaratively. A list comprehension takes the form:
\begin{tabbing}
aa \= aaa \= aaa \= aaa \= aaa \= aaa \= aaa \kill
\> \> {\tt [$T$ : $E_1$ {\tt in} $D_1$, $\ldots$, $E_n$ in $D_n$, $LocalVars$,$Goal$]} 
\end{tabbing}
where $LocalVars$ (optional) specifies a list of local variables, $Goal$ (optional) is a callable term. This construct means that for each permutation of values $E_1 \in D_1$, $\ldots$, $E_n \in D_n$, if the instance of $Goal$ with renamed local variables is true, then $T$ is added into the list. 
A list of this form is interpreted as a list comprehension if it occurs as an argument of a call to \verb+'@='/2+ or in arithmetic constraints.

Calls to {\tt foreach} and list comprehensions are translated into tail-recursive predicates. For example, the call {\tt Xs @= [X : (X,\_) in Ps]} is translated into
\begin{verbatim}
    dummy(Ps, L, []), Xs @= L
\end{verbatim}
where {\tt dummy} is defined with matching clauses as follows:
\begin{verbatim}
    dummy([],Xs,XsR) => Xs=XsR.
    dummy([(X,_)|Ps],Xs,XsR) => Xs=[X|Xs1], dummy(Ps,Xs1,XsR).
    dummy([_|Ps],Xs,XsR) => dummy(Ps,Xs,XsR).
    dummy(Ps,_,_) => throw(illegal_argument_in_foreach(Ps)).
\end{verbatim}
As can be seen in this example, there is little or no penalty to using these loop constructs compared with using recursion.

The loop constructs considerably enhance the modeling power of CLP(FD). The following gives two programs for the N-queens problem to illustrate different uses of the loop constructs. Here is the first program:
\begin{verbatim}
    queens(N,Qs):-
        length(Qs,N),
        Qs in 1..N,
        foreach(I in 1..N-1, J in I+1..N,
                (Qs[I] #\= Qs[J],
                 abs(Qs[I]-Qs[J]) #\= J-I)).
\end{verbatim}
The call {\tt queens(N,Qs)} creates a list {\tt Qs} of {\tt N} variables (one variable for each column), declares the domain of each of the variables to be {\tt 1..N}, and generates constraints to ensure that no two queens are placed in the same row or the same diagonal. 

The following program models the problem with Boolean constraints.
\begin{verbatim}
bool_queens(N,Qs):-
    new_array(Qs,[N,N]),
    Vars @= [Qs[I,J] : I in 1..N, J in 1..N],
    Vars in 0..1,
    foreach(I in 1..N,     % one queen in each row
            sum([Qs[I,J] : J in 1..N]) #= 1),
    foreach(J in 1..N,     % one queen in each column
            sum([Qs[I,J] : I in 1..N]) #= 1),
    foreach(K in 1-N..N-1, % at most one queen in each left-down diag
            sum([Qs[I,J] : I in 1..N, J in 1..N, I-J=:=K]) #=< 1),
    foreach(K in 2..2*N,   % at most one queen in each left-up diag
            sum([Qs[I,J] : I in 1..N, J in 1..N, I+J=:=K]) #=< 1).
\end{verbatim}
The call {\tt Vars @= [Qs[I,J] : I in 1..N, J in 1..N]} extracts the variables from matrix {\tt Qs} into list {\tt Vars}. List comprehensions are used in aggregate constraints. For example, the constraint \verb+sum([Qs[I,J] : J in 1..N]) #= 1+ means that the sum of the {\tt Ith} row of the matrix is equal to 1.

The {\tt foreach} construct of B-Prolog is different from the loop constructs provided by ECL$^i$PS$^e$ \cite{Schimpf02}. Syntactically, {\tt foreach} in B-Prolog is a variable-length call in which only one type of iterator, namely {\tt $E$ in $D$}, is used for iteration, and an extra argument is used for accumulators if needed. In contrast, ECL$^i$PS$^e$ provides a built-in, called {\tt do/2}, and a base iterator, named {\tt fromto/4}, from which six types of iterators are derived for describing various kinds of iteration and accumulation. In addition, in B-Prolog variables in a loop are assumed to be global unless they are declared local or occur in the patterns of the iterators ({\it global-by-default}). In contrast, in ECL$^i$PS$^e$ variables are assumed to be local unless they are declared global ({\it local-by-default}). From the programmer's perspective, the necessity of declaring variables is a burden in both approaches and no approach is uniformly better than the other. Nevertheless, small loops tend to have fewer local variables than global ones, and for them global-by-default tends to impose less a burden than local-by-default. For example, while the two N-queens programs shown above contain no declaration of local variables, in ECL$^i$PS$^e$ the variables {\tt N} and {\tt Qs} would have to be declared global. Large loop bodies, however, may require declaration of more local variables than global ones, but my personal opinion is that large loop bodies should be put in separate predicates for better readability. From the implementation perspective, ECL$^i$PS$^e$'s local-by-default can be easily implemented by goal expansion while B-Prolog's global-by-default requires analysis of variable scopes. B-Prolog issues warnings for occurrences in loop goals of singleton variables including anonymous variables.
 
Semantically, B-Prolog's iterators are matching-based while ECL$^i$PS$^e$'s iterators are unification-based. In B-Prolog, iterators never change collections unless the goal of the loop changes them explicitly. In contrast, in ECL$^i$PS$^e$ variables in collections can be changed during iterations even if the goal does not touch on the variables. This implicit change of variables in collections may make loops less readable.

\subsection{Tabling}
Tabling has been found increasingly important not only for helping beginners write workable declarative programs but also for developing real-world applications such as natural language processing, model checking, and machine learning applications. B-Prolog implements a tabling mechanism, called linear tabling \cite{Zhou08tab}, which is based on iterative computation of looping subgoals rather than suspension of them to compute the fixed points. The PRISM system \cite{sato01}, which heavily relies on tabling, has been the main driving force for the design and implementation of B-Prolog's tabling system.

The idea of tabling is to memorize the answers to tabled calls and use the answers to resolve subsequent variant calls. In B-Prolog, as in XSB, tabled predicates are declared explicitly by declarations in the following form:
\begin{tabbing}
aa \= aaa \= aaa \= aaa \= aaa \= aaa \= aaa \kill
\> \> {\tt :-table} $P_1$/$N_1$,$\ldots$,$P_k$/$N_k$.
\end{tabbing}
For example, the following tabled predicate defines the transitive closure of a relation as given by {\tt edge/2}.
\begin{verbatim}
    :-table path/2.
    path(X,Y):-edge(X,Y).
    path(X,Y):-path(X,Z),edge(Z,Y).
\end{verbatim}
With tabling, any query to the program is guaranteed to terminate as long as the term sizes are bounded.

By default, all the arguments of a tabled call are used in variant checking and all answers are tabled for a tabled predicate. B-Prolog supports table modes, which allow the system to use only input arguments in variant checking and table answers selectively. The table mode declaration
\begin{verbatim}
      :-table p(M1,...,Mn):C.
\end{verbatim}
\index{table declaration}
directs the system on how to do tabling on {\tt p/n}, where {\tt C}, called a {\it cardinality limit}, is an integer which limits the number of answers to be tabled, and each {\tt Mi} is a mode which can be {\tt min}, {\tt max}, {\tt +} (input), or {\tt -} (output). An argument with the mode {\tt min} or {\tt max}, called {\it optimized}, is assumed to be output. If the cardinality limit {\tt C} is 1, it can be omitted with the preceding ':'. In the current implementation, only one argument can be optimized. Since an optimized argument is not required to be numeral and the built-in {\verb+@<+/2} is used to select answers with minimum or maximum values, multiple values can be optimized.

The system uses only input arguments in variant checking, disregarding all output arguments. After an answer is produced, the system tables it unconditionally if the cardinality limit is not yet reached. When the cardinality limit has been reached, however, the system tables the answer only if it is better than some existing answer in terms of the argument with the {\tt min} or {\tt max} mode. In this case, the new answer replaces the worst answer in the table.

Mode-directed tabling in B-Prolog was motivated by the need to scale up the PRISM system\cite{sato01,Sato09,Zhou10tai} for handling large data sets. For a given set of possibly incomplete observed data, PRISM collects all explanations for these data using tabling and estimates the probability distributions by conducting EM learning~\cite{Dempster77} on these explanations. For many real-world applications, the set of explanations may be too large to be completely collected even in compressed form. Mode-directed tabling allows for collecting a subset of explanations. 

Mode-directed tabling is in general very useful for declarative description of dynamic programming problems \cite{GuoG08}. For example, the following program encodes Dijkstra's algorithm for finding a path with the minimum weight between a pair of nodes.
\noindent
\begin{verbatim}
    :-table sp(+,+,-,min).
    sp(X,Y,[(X,Y)],W) :-
        edge(X,Y,W).
    sp(X,Y,[(X,Z)|Path],W) :-
        edge(X,Z,W1),
        sp(Z,Y,Path,W2),
        W is W1+W2.
\end{verbatim}
The table mode states that only one path with the minimum weight is tabled for each pair of nodes. 

\subsection{Other Extensions and Features}

{\it The JIPL interface with Java:} This interface was designed and implemented by Nobukuni Kino, originally for his K-Prolog system, and was ported to B-Prolog. This bi-directional interface makes it possible for Java applications to use Prolog features such as search and constraint solving, and for Prolog applications to use Java resources such as networking, GUI, database, and concurrent programming packages. 

{\it PRISM \cite{Sato09}:} This is an extension of Prolog that integrates logic programming, probabilistic reasoning, and EM learning. It allows for the description of independent probabilistic choices and their logical consequences in general logic programs. PRISM supports parameter learning. For a given set of possibly incomplete observed data, PRISM can estimate the probability distributions to best explain the data. This power is suitable for applications such as learning parameters of stochastic grammars, training stochastic models for gene sequence analysis, game record analysis, user modeling, and obtaining probabilistic information for tuning systems performance. PRISM offers incomparable flexibility compared with specific statistical model such as Hidden Markov Models (HMMs), Probabilistic Context Free Grammars (PCFGs) and discrete Bayesian networks. PRISM is a product of the PRISM team at Tokyo Institute of Technology led by Taisuke Sato.

{\it CGLIB \cite{Zhou03:cglib}:} This is a constraint-based high-level graphics library developed for B-Prolog. It supports over twenty types of basic graphical objects and provides a set of constraints including non-overlap, grid, table, and tree constraints that facilitates the specification of layouts of objects. The constraint solver of B-Prolog serves as a general-purpose and efficient layout manager, which is significantly more flexible than the special-purpose layout managers used in Java. The library uses action rules available in B-Prolog for creating agents and programming interactions among agents or between agents and users. CGLIB is supported in the Windows version only.

{\it Logtalk \cite{Moura09}:} This is an extension of Prolog developed by Paulo Moura that supports object-oriented programming. It runs with several Prolog systems. Thanks to Paulo Moura's effort, Logtalk has been made to run with B-Prolog seamlessly. Logtalk can be used as a module system on top of B-Prolog. 

{\it The LP/MIP interface:} B-Prolog provides an interface to LP/MIP (linear programming and mixed integer programming) packages such as GLPK and CPLEX. With the declarative loop constructs, B-Prolog can serve as a powerful modeling language for LP/MIP problems.

\section{The TOAM Jr. Prolog Machine}
We assume that all the clauses of a given Prolog program have been translated into matching clauses where input and output unifications are separated and the determinacy is denoted explicitly. This form of matching clauses is called {\tt canonical form}. Consider, for example, the {\tt append} predicate in Prolog:

\clearpage
\begin{verbatim}
     append([],Ys,Ys).
     append([X|Xs],Ys,[X|Zs]):-
        append(Xs,Ys,Zs).
\end{verbatim}
This program is translated equivalently into the following matching clauses with no assumption on modes of arguments:

  \begin{minipage}[t]{.45\textwidth}
\begin{verbatim}
   append(Xs,Ys,Zs),var(Xs) => 
      append_aux(Xs,Ys,Zs).
   append([],Ys,Zs) =>  
      Ys=Zs.
   append([X|Xs],Ys,Zs) =>
      Zs=[X|Zs1],
      append(Xs,Ys,Zs1).

\end{verbatim}
  \end{minipage}
  \hfill
  \begin{minipage}[t]{.45\textwidth}
\begin{verbatim}
append_aux(Xs,Ys,Zs) ?=> 
   Xs=[],
   Ys=Zs.
append_aux(Xs,Ys,Zs) =>
   Xs=[X|Xs1],
   Zs=[X|Zs1],
   append(Xs1,Ys,Zs1).
\end{verbatim}
  \end{minipage}

\noindent
The B-Prolog compiler does not infer modes but makes use of modes supplied by the programmer to generate more compact canonical-form programs. For example, with the mode declaration
\begin{verbatim}
   :-mode append(+,+,-).
\end{verbatim}
The append predicate is translated into the following canonical form:
\begin{verbatim}
   append([],Ys,Zs) =>  
      Ys=Zs.
   append([X|Xs],Ys,Zs) =>
      Zs=[X|Zs1],
      append(Xs,Ys,Zs1).
\end{verbatim}
The compiler does not check modes at compile time or generate code for verifying modes at runtime.

\subsection{The Memory Architecture}
Except for changes made to accommodate event handling and garbage collection as to be detailed below, the memory architecture is the same as the ATOAM \cite{Zhou96} employed in early versions of B-Prolog. 

\begin{center}
\begin{figure}[h]
\epsfxsize=5cm 
\centering{\epsfbox{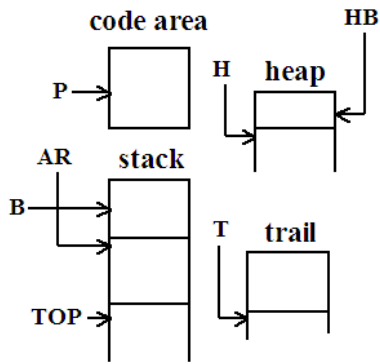}}
\caption{\label{fig:mem}The Memory Architecture of TOAM Jr.}
\end{figure}
\vspace*{-1cm}
\end{center}

\subsubsection{Code and data areas}
TOAM Jr. uses all the stacks and data areas used by the WAM (see Figure \ref{fig:mem}). There is a data area called {\it code area} that contains, besides instructions compiled from programs,  a symbol table that stores information about the atom, function, and predicate symbols in the programs. There is one record for each symbol in the table which stores such information as the {\it name}, {\it arity}, {\it type}, and {\it entry point} if the symbol is defined.  

The {\it stack} stores frames associated with predicate calls. Predicate call arguments are passed through the stack and only one frame is used for each predicate call. Each time a predicate is called, a frame is placed on top of the stack unless the frame currently at the top can be reused. Frames for different types of predicates have different structures. For standard Prolog, a frame is either {\it determinate} or {\it nondeterminate}. A nondeterminate frame is also called a {\it choice point}. 

The {\it heap} stores terms created during execution. 

The {\it trail} stack stores updates that must be undone upon backtracking. The use of a trail stack to support backtracking is the major difference between Prolog machines and machines for other languages such as Pascal, Lisp, and Java.

\subsubsection{Term representation}
Terms are represented in the same way as in the WAM \cite{Warren83}. A term is represented by a word containing a value and a tag. The tag
distinguishes the type of the term. It may be {\tt REF} denoting a reference, {\tt INT} denoting an integer, {\tt ATM} denoting an atom,
{\tt STR} denoting a structure, or {\tt LST} denoting a cons. A float is treated as a special structure in our implementation. Another tag is used for suspension variables including domain variables.

\ignore{
The value of a term is an address except when the term is an integer
(in this case, the value represents the integer itself). The address
points to a different place depending on the type of the term. A {\it free}
variable is represented by a self-referencing pointer. A free variable stored on the stack is called a {\it stack variable} and a free variable stored on the heap is called a {\it heap variable}. The operation that looks for the value at the end of a reference chain is called {\it dereference}.  The address in an atom points to the record for the atom in the symbol table. The address in a structure $f(t_1,\ldots,t_n)$ points to a block of $n+1$ consecutive words on the heap where the first word points to the record for the functor $f/n$ in the symbol table, and the remaining $n$ words store the $n$ arguments of the structure. The address in a cons $[H|T]$ points to a block of two consecutive words in the heap where the first word stores the head $H$, and the second word stores the tail $T$. As in Lisp implementations, this special representation for lists benefits applications where lists are heavily used.

For example, Figure \ref{fig:term} shows a representation of the term {\tt f(X,a,X,[1])}, where {\sf self} is a self-referencing pointer.

\begin{figure}[h]
\vspace*{-5mm}
\begin{center}
\epsfxsize=5cm 
\centering{\epsfbox{term.eps}}
\caption{\label{fig:term}A representation of {\tt f(X,a,X,[1]).}}
\end{center}
\vspace*{-5mm}
\end{figure}
}

\subsubsection{Registers}
The following registers are used to represent the machine status (see Figure \ref{fig:mem}):

\begin{tabular}{ll} \hline
{\tt P:} & Current program pointer \\
{\tt TOP:} & Top of the stack \\
{\tt AR:}  & Current frame pointer \\
{\tt H:} & Top of the heap \\
{\tt T:} & Top of the trail stack \\
{\tt B:} & Latest choice point \\
{\tt HB:} & {\tt H} slot of the latest choice point, i.e., {\tt B->H} \\ \hline
\end{tabular}
The {\tt HB} register, which also exists in the WAM, is an alias for {\tt B->H}. It is used in checking whether or not a variable needs to be trailed. When a free variable is bound, if it is a heap variable older than {\tt HB} or a stack variable older than {\tt B}, then it is trailed.

\subsubsection{Stack frame structures}
Frames for different types of predicates have different structures. A determinate frame has the following structure:

\begin{tabular}{ll} \hline
{\tt A1..An}: &  Arguments \\
{\tt AR}:     &  Parent frame pointer\\
{\tt CP}:     &  Continuation program pointer \\
{\tt BTM}:    &  Bottom of the frame \\
{\tt TOP}:    &  Top of the frame \\
{\tt Y1..Ym}: & Local variables \\ \hline
\end{tabular}

\noindent
Where {\tt BTM} points to the bottom of the frame, i.e., the slot for the first argument {\tt A1}, and {\tt TOP} points to the top of the frame, i.e., the slot just next to that for the last local variable {\tt Ym}. The {\tt BTM} slot was not in the original ATOAM \cite{Zhou96}. This slot was introduced to support garbage collection and event-driven action rules which require a new type of frames called {\it suspension frames} \cite{Zhou06ar}. The {\tt AR} register points to the {\tt AR} slot of the current frame. Arguments and local variables are accessed through offsets with respect to the {\tt AR} slot. 

It is the caller's job to place the arguments and fill in the {\tt AR} and {\tt CP} slots. The callee fills in the {\tt BTM} and {\tt TOP} slots.

A choice point contains, besides the slots in a determinate frame, four slots located between the {\tt TOP} slot and local variables: 

\begin{tabular}{ll} \hline
{\tt CPF}: &  Backtracking program pointer \\
{\tt H}:   &  Top of the heap \\
{\tt T}:   &  Top of the trail \\
{\tt B}:   &  Previous choice point \\ \hline
\end{tabular}

\noindent
The {\tt CPF} slot stores the program pointer to continue with when the current branch fails. The slot {\tt H} points to the top of the heap  and {\tt T} points to the top of the trail stack when the frame was allocated. When a variable is bound, it must be trailed if it is older than {\tt B} or {\tt HB}. When execution backtracks to the latest choice point, the bound variables trailed on the trail stack between {\tt T} and {\tt B->T} are set back to free, the machine status registers {\tt H} and {\tt T} are restored, and the program pointer {\tt P} is set to be {\tt B->CPF}. 

The original ATOAM presented in \cite{Zhou96} had another type of frame, called {\it non-flat}, for determinate predicates that have non-flat or deep guards. This frame was abandoned since it is difficult for the compiler to extract non-flat guards to take advantage of this feature. 

\subsubsection{\label{sec:assertions}Assertions}
The following assertions always hold during execution:
\begin{enumerate}
\item No heap cell can reference a stack slot.
\item No older stack slot can reference a younger stack slot and no older heap variable can reference a younger heap variable.
\item No slot in a frame can reference another slot in the same frame.
\end{enumerate}
Assertions 1 and 2 are also enforced by the WAM. The third assertion is needed to make dereferencing the arguments of a last call unnecessary when the current frame is reused. To enforce this assertion, when two terms being unified are stack variables, the unification procedure {\it globalizes} them by creating a new heap variable and letting both stack variables reference it.

\subsection{The base instruction set}
Figure \ref{fig:base} gives TOAM Jr.'s base instruction set. An instruction with operands is denoted as a Prolog structure whose functor denotes the name and whose arguments denote the operands; an instruction with no operand is denoted as an atom. An operand is either a frame slot $y$, an integer literal $i$, a label $l$, a constant $a$, a functor $f/n$, a predicate symbol $p/n$, or a {\it tagged} operand $z$. If an instruction carries two or more operands of the same type, subscripts are used to differentiate them. 

In the examples below, an operand $y$ occurs as {\tt y(I)} where {\tt I} is the offset w.r.t. the {\tt AR} slot of the current frame,\footnote{Arguments have positive offsets and local variables have negative offsets.} and a tagged operand $z$ occurs as one of the following:
\begin{itemize}
\item $v(i)$ denotes an uninitialized frame slot with offset $i$. This is for the first occurrence of a variable. A singleton variable, also called a dummy variable, is denoted as $v(0)$.
\item $u(i)$ denotes an initialized frame slot with offset $i$. This is for subsequent occurrences of a variable.

\item $c(a)$ denotes a constant $a$.
\end{itemize}
To distinguish among these three cases, a tagged operand carries a {\em  code tag}. So if an operand is $v(i$), it occurs in the instruction as the integer $i$ with a tag; if it is $u(i)$, it occurs as the integer $i$ with a different tag; and if it is $c(a)$, it occurs as the WAM representation of $a$, i.e., an {\tt INT} tagged integer or an {\tt ATM} tagged atom.

\begin{figure}[bt]
\begin{minipage}[t]{.55\textwidth}
\begin{tabbing}
aa \= aaa \= aaa \= aaa \= aaa \= aaa \= aaa \kill
Control: \\
\> {\tt allocate\_det}($i_1$,$i_2$) \\
\> {\tt allocate\_nondet}($i_1$,$i_2$) \\
\> {\tt return} \\
\> {\tt fork}($l$) \\
\> {\tt cut} \\
\> {\tt fail} \\
\\
Branch: \\
\> {\tt jmpn\_constant}($y,l_{var},l_{fail},a$) \\
\> {\tt jmpn\_struct}($y,l_{var},l_{fail},f/n,y_1,\ldots,y_n$) \\
\> {\tt switch\_on\_cons}($y,l_{nil},l_{var},l_{fail},y_1,y_2$) \\
\> {\tt hash}($y,i,(val_1,l_1),\ldots,(val_i,l_i),l_{var},l_{fail}$) \\
\end{tabbing}
\end{minipage}
\begin{minipage}[t]{.45\textwidth}
\begin{tabbing}
aa \= aaa \= aaa \= aaa \= aaa \= aaa \= aaa \kill
Unify: \\
\> {\tt unify\_constant}($y,a$) \\
\> {\tt unify\_value}($y_1,y_2$) \\
\> {\tt unify\_struct}($y,f/n,z_1,\ldots,z_n$) \\
\> {\tt unify\_list}($y,i,z_1,\ldots,z_i,z_{i+1}$) \\
\\
Move: \\
\> {\tt move\_struct}($y,f/n,z_1,\ldots,z_n$) \\
\> {\tt move\_list}($y,i,z_1,\ldots,z_i,z_{i+1}$) \\
\\
Call: \\
\> {\tt call}($p/n,z_1,\ldots,z_n$) \\
\> {\tt last\_call}$(i,p/n,z_1,\ldots,z_n$) 
\end{tabbing}
\end{minipage}
\caption{\label{fig:base} The TOAM Jr. base instruction set.}
\vspace*{-3mm}
\end{figure}

\subsubsection{Control instructions}
The first instruction in the compiled code of a predicate is an allocate instruction which takes two operands: the arity and the size of the frame, not counting the arguments. By the time an allocate instruction is executed, the arguments of the current call should have been placed on top of the stack, and the {\tt AR} and {\tt CP} slots should have been set by the caller. An allocate instruction is responsible for fixing the size of the current frame and saving status registers if necessary. In the actual implementation, an allocate instruction also handles events and interrupt signals if there are any. For the sake of simplicity, these operations are not included in the definition. Nevertheless, it is assumed that any predicate can be interrupted and preempted by event handlers. Therefore, a runtime test is needed in {\tt last\_call} to determine if the current frame can be deallocated or reused.

\begin{itemize}
\item The {\tt allocate\_det} instruction starts the code of a determinate predicate. It sets the {\tt BTM} and {\tt TOP} slots and updates the {\tt TOP} register.
\ignore{
\begin{tabbing}
aa \= aaa \= aaa \= aaa \= aaa \= aaa \= aaa \kill
\>   {\tt allocate\_det(arity,size)\{} \\
\> \>    {\tt AR->BTM = AR+arity;} \\
\> \>    {\tt TOP = AR-size;} \\
\> \>    {\tt AR->TOP = TOP;} \\
\>   {\tt \}}
\end{tabbing}
}
\item The {\tt allocate\_nondet} instruction starts the code of a nondeterminate predicate. In addition to fixing the size of the frame, it also saves the contents of the status registers into the frame.
\ignore{
\begin{tabbing}
aa \= aaa \= aaa \= aaa \= aaa \= aaa \= aaa \kill
\>   {\tt allocate\_nondet(arity,size)\{} \\
\>\>      {\tt allocate\_det(arity,size);} \\
\>\>      {\tt AR->B = B;} \\
\>\>      {\tt AR->H = H;} \\
\>\>      {\tt AR->T = T;} \\
\>\>      {\tt HB = H;} \\
\>   {\tt \}}
\end{tabbing}
}
\item The {\tt return} instruction returns control to the caller, and deallocates the frame if the current frame is the topmost one that is not pointed to by the {\tt B} register.
\ignore{
\begin{tabbing}
aa \= aaa \= aaa \= aaa \= aaa \= aaa \= aaa \kill
\>   {\tt return()\{ } \\
\>\>      {\tt P = AR->CP;} \\ 
\>\>      {\tt if (B!=AR \&\& AR->TOP==TOP) TOP = AR->BTM;} \\
\>\>      {\tt AR = AR->AR;} \\  
\>   {\tt \} }
\end{tabbing}
}
In ATOAM, when the current frame is deallocated, the top of the stack is set to the top of the parent frame or the latest choice point, whichever is younger. With event handling, however, this becomes unsafe because the chain of active frames, forming a spaghetti stack, are not in chronological order \cite{Zhou06ar}. For this reason, the top of the stack is set to the bottom of the current frame after it is deallocated.

\item The {\tt fork} instruction sets the {\tt CPF} slot of the current frame.
\ignore{
\begin{tabbing}
aa \= aaa \= aaa \= aaa \= aaa \= aaa \= aaa \kill
\>   {\tt fork(addr)\{} \\
\>\>      {\tt AR->CPF = addr;} \\
\>   {\tt \}}
\end{tabbing}
}
\item The {\tt cut} instruction discards the alternative branches of this frame and all the choice points that are younger.
\item The {\tt fail} instruction lets execution backtrack to the latest choice point.
\end{itemize}

\subsubsection*{Example}
The following shows a canonical-form program and its compiled code:
\begin{verbatim}
   %    p ?=> true.
   %    p => true.
   p/0: allocate_nondet(0,8)
        fork(l1)
        return
    l1: cut
        return
\end{verbatim}
Since the predicate is nondeterminate and there is no local variable, the allocated frame contains 8 slots reserved for saving the machine status.

\subsubsection{Branch instructions}
Unification calls in the guards of clauses in a predicate are encoded as {\it branch} instructions. Each branch instruction takes a label $l_{fail}$ to go to on failure of the test and also a label $l_{var}$ to go to when the tested operand is a variable. The {\tt jmpn\_struct} instruction fetches the arguments of the tested structure into designated frame slots when the test is successful. The {\tt switch\_on\_cons} instruction moves control to the next instruction if the tested operand is a cons and to $l_{nil}$ if it is an empty list. When the tested operand is a cons, the instruction also fetches the head and tail of the cons into the designated frame slots. The {\tt hash} instruction determines the address of the next instruction based on the tested operand and a hash table.

The following shows an example.
\begin{verbatim}
   %    p(F),F=f(A),A=a => true.
   p/1: allocate_det(1,4)
        jmpn_struct(y(1),l_fail,l_fail,f/1,y(1))
        jmpn_constant(y(1),l_fail,l_fail,a)
        return
\end{verbatim}
The operand {\tt l\_fail} is the address of a {\tt fail} instruction. Notice that the argument slot with offset 1 allocated to the variable {\tt F} is reused for {\tt A}.  None of the branch instructions carries tagged operands.

\subsubsection{Unify instructions}
Recall that in canonical-form Prolog every unification call in the bodies takes the form $V=T$ where $V$ is a variable and $T$ is either a variable, a constant, a list with no compound elements, or a compound term with no compound arguments. A unify instruction encodes a unification call where neither $V$ nor $T$ is a first-occurrence variable in the clause. For each type of $T$, there is a type of unify instruction. The {\tt unify\_constant} instruction is used if $T$ is a constant; {\tt unify\_value} is used if $T$ is a variable; {\tt unify\_list} is used if $T$ is a list, and {\tt unify\_struct} is used if $T$ is a structure. The {\tt unify\_list}$(y,i,z_1,\ldots,z_i,z_{i+1})$ encodes the list [$z_1,\ldots,z_i$$|$$z_{i+1}$]. The {\tt unify\_struct} and {\tt unify\_list} instructions have variable lengths and the operands for list elements or structure arguments are all tagged. In a {\tt unify\_struct} instruction, the number of tagged operands is determined by the functor $f/n$; and in a {\tt unify\_list} instruction the number is given as a separate operand.

A unify instruction for $V=T$ unifies the term referenced by $V$ with $T$ if $V$ is not free. In WAM's terminology, the unification is said to be in {\tt read} mode in this case. If $V$ is a free variable, the instruction builds the term $T$ and binds $V$ to the term. This mode is called {\tt write} in the WAM. Since a unification is encoded as only one instruction, there is no need to use a register for the mode.

Special care must be taken to ensure that no heap cell references a stack slot. The {\tt unify\_list} and {\tt unify\_struct} instructions must dereference a tagged operand if the operand is not a first-occurrence variable and globalize it if the dereferenced term is a stack variable. 
\ignore{
This dereference operation, however, is not as expensive as the general dereference operation since it stops walking the chain once the content of a stack slot is found to be a reference to the heap.
}

The following shows an example.
\begin{verbatim}
   %    p(F) => F=f(L),L=[X,X,a].
   p/1: allocate_det(1,4)
        unify_struct(y(1),f/1,v(1))
        unify_list(y(1),3,v(1),u(1),c(a),c([])
        return
\end{verbatim}
The argument slot with offset 1 allocated to the variable {\tt F} is reused for {\tt L} and later also for {\tt X}. Since {\tt L} is a first-occurrence variable, it is encoded as the tagged operand {\tt v(1)}. The variable {\tt X} occurs twice in {\tt L=[X,X,a]}. The first occurrence is encoded as {\tt v(1)} and the second one is encoded as {\tt u(1)}. The tagged operand {\tt c(a)} encodes the constant element {\tt a} and the operand {\tt c([])} encodes the empty tail of the list.

\subsubsection{Move instructions}
A move instruction is used to encode a unification $V=T$ where $V$ is a first-occurrence variable in the clause. $T$ is assumed to be a compound term. If $T$ is a constant or a variable, the unification can be performed at compile time by substituting all occurrences of $V$ for $T$ in the clause. For this reason, only {\tt move\_struct} and {\tt move\_list} instructions are needed.

\subsubsection{Call instructions}
A {\tt call} instruction encodes a non-last call in the body of a clause.
\begin{tabbing}
aa \= aaa \= aaa \= aaa \= aaa \= aaa \= aaa \kill
\>   {\tt call}($p/n,z_1,\ldots,z_n$){\tt \{} \\
\>\>       for each $z_i$ ($i=1,...,n$) do \\
\>\>\>         {\tt *TOP-- = }value of $z_i$ \\
\>\>      {\tt parent\_ar = AR;} \\
\>\>      {\tt AR = TOP;} \\
\>\>      {\tt AR->AR = parent\_ar;} \\
\>\>      {\tt AR->CP = P;} \\
\>\>      {\tt P = entrypoint($p/n$);} \\
\>   {\tt \}}
\end{tabbing}
After passing the arguments to the callee's frame, the instruction also sets the {\tt AR} and {\tt CP} slots of the frame, and lets the {\tt AR} register point to the frame.

The value of each tagged operand $z_i$ is computed as follows. If it is $v(k)$, then the value is the address of the frame slot with offset $k$ (it is initialized to be a free variable) unless when $k$ is $0$, in which case the value is the content of the {\tt TOP} register. If it is $u(k)$, then the value is the content of the frame slot with offset $k$. Otherwise, the value is $z_i$ itself, which is a tagged constant.

A {\tt last\_call} instruction encodes the last call in the body of a determinate clause or a clause in a nondeterminate predicate that contains cuts. For a nondeterminate clause in a nondeterminate predicate that does not contain cuts, the last call is encoded as a {\tt call} instruction followed by a {\tt return} instruction. Unlike the {\tt call} instruction which always allocates a new frame for the callee, the {\tt last\_call} instruction reuses the current frame if it is a determinate frame or a choice point frame whose alternatives have been cut off. The {\tt last\_call} instruction takes an integer, called {\it layout bit vector}, which tells what arguments are misplaced and hence need to be rearranged into proper slots in the callee's frame when the current frame is reused. There is a bit for each argument and the argument needs to be rearranged if its bit is 1.\footnote{In the actual implementation, an integer with 28 bits is used for a layout vector. If the last call has more than 28 arguments, then the last-call optimization is abandoned.}

\begin{tabbing}
aa \= aaa \= aaa \= aaa \= aaa \= aaa \= aaa \kill
  {\tt last\_call(layout,$p/n,z_1,\ldots,z_n$)\{} \\
\>       {\tt if (AR->TOP==TOP \&\& B$!=$AR)\{\ /* reuse */} \\
\>\>        {\tt for each argument $z_i (i=1,\ldots,n)$ do } \\
\>\>\>            {\tt if ($z_i$ is tagged u and its layout bit is 1)} \\
\>\>\>  \>           {\tt copy $z_i$ to a temporary frame;} \\
\>\>       {\tt move AR->AR and AR->CP if necessary;} \\
\>\>       {\tt arg\_ptr = AR->BTM+1;} \\
\>\>       {\tt for each argument $z_i (i=1,\ldots,n)$ do} \\
\>\>\>          {\tt if ($z_i$'s layout bit is 1)} \\
\>\>\> \>           {\tt *(arg\_ptr-i) = the value $z_i$;} \\
\>\>      {\tt AR = AR+(AR->BTM)-n; } \\
\>\>      {\tt P = entrypoint($p/n$);} \\
\>      {\tt \} else} \\
\>\> {\tt call($p/n,z_1,\ldots,z_n$);} \\
 {\tt \}}
\end{tabbing}
The following steps are taken to reuse the current frame: Firstly, all the misplaced arguments that are tagged {\tt u} are copied out to a temporary frame. Because of the enforcement of assertion 3, it is unnecessary to fully dereference stack slots, but free variables in the frame must be globalized since otherwise unrelated arguments may be wrongly aliased. Constants and first-occurrence variables in the arguments are not touched in this step. Secondly, if the arity of the current frame is different from the arity of the last call, the {\tt AR} and {\tt CP} slots are moved. Thirdly, all misplaced arguments are moved into the frame for the callee. For {\tt u}-tagged arguments, the values in the temporary frame are used instead of the old ones because the old values may have been overwritten by other values. Finally, the {\tt AR} register is set to be {\tt AR+(AR->BTM)-n}.

For example,
\begin{verbatim}
   % p(X,Y,Z) => S=f(X,Y),q(S),r(Z,Y,X,W).
   p/3: allocate_det(3,5)
        move_struct(y(-1),f/2,u(3),u(2))  % S=f(X,Y)
        call(q/1,u(-1))                   % q(S)
        last_call(0b1011,r/4,u(1),u(2),u(3),v(0))
\end{verbatim}
The binary literal '0b1101' is the layout bit vector for the last call which indicates that all the arguments except for the second one ({\tt Y}) are misplaced. The variable {\tt W} is a singleton variable in the clause and is encoded as {\tt v(0)}.

\subsection{Storage allocation}
Each variable is allocated a frame slot and is accessed through the offset of the slot. All singleton variables have offset 0. When an operand is tagged $v$, the offset must be tested. If the offset is 0, then it is known to be a singleton variable.\footnote{A variable with offset 0 is never stored in the current frame. Recall that the slot with offset 0 stores the pointer to the parent frame.}

Frame slots allocated to variables are reclaimed as early as possible such that they can be reused for other variables. A variable is said to be {\it inactive} if it is not accessible in either forward or backward execution. The storage allocated to a variable can be reclaimed immediately after the call in which the variable becomes inactive. Because of the existence of nondeterminate predicates, a variable may still be active even after its last occurrence. For example, consider the clause
\begin{verbatim}
  a(U) => b(U,V),c(V,W),d(W).
\end{verbatim}
The slot allocated to {\tt U} can be reused after {\tt b(U,V)} since the clause is determinate, but if {\tt b/2} is nondeterminate the slot allocated to {\tt V} cannot be reused even after {\tt c(V,W)}.

\subsection{Instruction specialization and merging}
Instruction specialization and merging are two well-known important techniques used in abstract machine implementations. The omission of registers can make these techniques more effective. In this section, we discuss how some base instructions are specialized and what instructions are merged.

\subsubsection{Instruction specialization}
The variable length instructions that take tagged operands are targets for specialization. A variable length instruction is more expensive to interpret than a fixed length instruction since the emulator needs to fetch the number of operands and iterate through the operands using a loop statement. A tagged operand is more expensive to interpret than an untagged one because its interpretation involves the following overhead: (1) testing the tag; (2) untagging the operand if it is a variable tagged $u$ or $v$; and (3) testing if the offset is 0 if the operand is an uninitialized variable tagged $v$. For an instruction of length up to $n$, $\Sigma_{i=1}^{n}3^i$ specialized instructions can be created (recall that there are three different operand types, namely, $v(i)$, $u(i)$, and $c(a)$). Obviously, reckless introduction of specialized instructions will result in explosion of the emulator size and even performance degradation depending on the platform. 

A specialized instruction carries the number and the types of its operands in its opcode. An instruction, named {\tt unify\_cons($y,z_1,z_2$)}, is introduced to replace {\tt unify\_list} that has two operands. The {\tt unify\_cons} instruction is further specialized so no operand is tagged.

Specialized instructions are introduced for {\tt unify\_struct} that has up to two arguments so no operand is tagged. Any {\tt unify\_struct} instruction that has more than two arguments is translated to a specialized instruction for the first two arguments followed by {\tt unify\_arg} instructions. In this way, no operand is tagged.

For the {\tt call} instruction, specialized instructions in the form of {\tt call\_$k$\_u} $(k=1,...,9)$ are introduced which carry $k$ initialized variables as operands in addition to the predicate symbol. Specialized instructions are also introduced for often-occurring call patterns such as {\tt u}, {\tt v}, and {\tt uv}.

Specialized versions of the {\tt last\_call} instruction are introduced that carry indices of misplaced arguments explicitly as operands. In general, a specialized instruction for a last call takes the form {\tt last\_call\_k($i_1,\ldots,i_k,p/n,z_1,\ldots,z_n)$} where the integers $i_1,\ldots,i_k$ are indices of misplaced arguments that need to be rearranged. The currently implemented abstract machine has three specialized instructions ($k=0,1,2$). Further specialized instructions are used to encode tail-recursive calls.

\subsubsection{Instruction merging}
The dispatching cost is considered one of the biggest sources of overhead in abstract machine emulators. Even with fast dispatching techniques such as threaded code, the overhead cannot be neglected. A widely used technique in abstract machine implementations for reducing the overhead is called instruction merging, which amounts to combining several instructions into one. Although our instructions have large granularity, there are still opportunities for merging instructions. 

It is often the case that a {\tt switch\_on\_cons} or {\tt fork} instruction is followed by a unify instruction and it is also often the case that a unify instruction is followed by a {\tt cut} instruction. So it makes sense to introduce merged instructions for these cases. We also introduce merged unify instructions for combining unify instructions and {\tt return}. In addition, {\tt cut} and {\tt fail} are merged as well as {\tt cut} and {\tt return}.

When merging two instructions, we do not just combine the routines for the original instructions to create the routine for the merged instruction. Sometimes the merged instruction can be interpreted more efficiently. For example, consider the merged instruction {\tt fork\_unify\_constant($l,y,a$)}, which combines {\tt fork($l$)} and {\tt unify\_constant($y,a$)}. The alternative program pointer {\tt CPF} is set to be $l$ if the unification succeeds. If the unification fails, however, execution can simply jump to $l$ because the machine status has  not changed since the creation of the frame. So the merged instruction not only saves the setting of {\tt CPF} but also replaces expensive backtracking with cheap jumping.

The same idea can be applied to merged instructions of unify and cut. Consider the merged instruction {\tt unify\_constant\_cut($y,a$)}. If $y$ is a free variable, then {\tt cut} can be performed before $y$ is bound to $a$. In this way, unnecessary trailing of $y$ can be avoided.

\subsection{Discussion}
Compiling a high-level language into an abstract or virtual machine has become a popular implementation method, which has traditionally been adopted by compilers for Lisp and Prolog, and recently made popular by implementations of Java and Microsoft .NET. One of the biggest issues in designing an abstract machine concerns whether to have arguments passed through registers or stack frames. Stack-based abstract machines are more common than register-based machines as exemplified by the Java Virtual Machine and Microsoft Intermediate Language.

One of the biggest advantages of passing arguments through stack frames over through registers is that instructions for procedure calls need not take destinations of arguments explicitly as operands. This leads to more compact bytecode and less interpretation overhead as well. For historical reasons, most Prolog systems are based on the WAM, which is a register machine, except for B-Prolog which is based on a stack machine called ATOAM. Even ATOAM retains registers for temporary variables.

For Prolog, a register machine such as the WAM does have its merits even when registers are normally simulated. Firstly, no frame needs to be created for determinate binary programs. Secondly, registers are represented as global variables in C and the addresses of the variables can be computed at load time rather than run time. Thirdly, in some implementations a register never references a stack slot, and hence when building a compound term on the heap the emulator needs not dereference a component if it is stored in a register. In a highly specialized abstract machine such as the one adopted in Quintus Prolog (according to \cite{NassenCS01}), the registers an instruction manipulates can be encoded as part of the opcode rather than taken explicitly as operands. In this way, if the emulator is implemented in an assembly language to which hardware registers are directly available, abstract machine registers can be mapped to native registers.

Nevertheless, using registers has more cons than pros for Prolog emulators. Firstly, as mentioned above, instructions for procedure calls have to carry destination registers as operands which results in less compact code. Secondly, long-lived data stored in registers have to be saved in stack frames and loaded later when they are used. In Prolog, variables shared by multiple chunks\footnote{A chunk consists of a non-inline call preceded by inline calls. The head of a clause belongs to the chunk of the first non-inline call in the body.} or multiple clauses are long-lived. 
\ignore{
According to the statistics reported by Zhou \cite{Zhou96}, passing arguments through registers only pays off when accessing register is five times faster than accessing memory.\footnote{That statistics did not take initialization of local stack variables into account.} In an abstract machine emulator where registers are simulated, this is never the case even when the addresses of ``registers'' are computed at load time. 
}
Thirdly, the information in a frame cannot be easily reused by the last call if the clause of the frame contains multiple chunks. Extra efforts are needed to reuse frames for such clauses \cite{DemoenN08a,Meier91}. Finally, registers make it more expensive to interpret tagged operands and harder to combine instructions because two more operand types, namely, uninitialized and initialized register variables, have to be considered. An alternative approach to facilitating instruction merging is to store all data in registers, as done in the BinWAM \cite{Tarau90}. Nevertheless, this approach causes overhead on non-binary clauses because of the necessity to create continuations as first-class terms.

\ignore{Finally, register machines are not suitable for event handling. B-Prolog is  undisputedly the winner now as far as constraint propagation speed is concerned \cite{Zhou06ar}.

The need to rearrange arguments of last calls is considered a weakness of ATOAM \cite{Demoen00b}. TOAM Jr. inherits the memory architecture of ATOAM as well as this necessity. In the WAM, arguments need to be rearranged into proper registers for first calls. It is easy to find program patterns that make one machine arbitrarily worse than the other.

 Nevertheless, our investigation of a large number of programs shows that last calls have more to share with the heads than first calls in most tail recursive predicates. The code generation for last calls for TOAM Jr. is much simpler than that for ATOAM. The ATOAM compiler adopts a sophisticated algorithm for generating code for last calls. For a last call, it builds a bipartite graph mapping the locations of arguments between the current and new frames, and optimizes the number of move instructions needed to rearrange the arguments. 
}
Another design issue of abstract machines concerns the granularity of instructions. The WAM has a fine-grained instruction set in the sense that an instruction roughly encodes a symbol in the source program. The ATOAM follows the WAM as far as granularity of instructions is concerned. There are Prolog machines that provide even more fine-grained instructions such as explicit dereference instructions \cite{vanroy94}. A fine-grained instruction set opens up more operations for optimization in a native compiler, but hinders fast interpretation due to a high dispatching cost. Instruction specialization and merging are two widely used techniques in abstract machine emulators for reducing the cost of interpretation \cite{Costa99,bart:wamvariations,NassenCS01}. In terms of granularity of instructions, TOAM Jr. resides between the WAM and a Prolog interpreter \cite{Maier88} where terms are interpreted without being flattened.  The use of coarse-grained instructions reduces the code size and the number of executed instructions for programs, leading to a reduced dispatching cost. 

Nevertheless, the interpretation of a variable number of tagged operands imposes certain overhead. After terms are flattened and registers are omitted, the number of possible operand types is reduced to three (constants, uninitialized variables and initialized variables). Because of the existence of only three operand types, the cost of interpreting tagged operands is smaller than the dispatching cost. Also, the specialization of most frequently executed instructions makes interpretation of tagged operands unnecessary.

\ignore{
Currently our compiler basically performs no program analysis and uses no information on modes of arguments or determinacy of predicate calls. Mode information is useful for translating a program into a more compact and efficient canonical form,  and determinacy information is useful to help the storage allocator detect the life spans of variables and allow variables to share frame slots. It is a future task to introduce a program analyzer to infer and make use of these kinds of useful information.
}

\section{Architectural Support for the Extensions}
This TOAM architecture has been extended to support action rules \cite{Zhou06ar} and tabling \cite{Zhou08tab}. This section overviews the changes to the memory architecture.

\subsection{Architectural support for action rules}
When a predicate defined by action rules is invoked by a call, a {\it suspension frame} is pushed onto the stack, which contains, besides the slots in a determinate frame, the following four slots:

\begin{tabular}{ll} \hline
{\tt STATE}: &     State of the frame \\
{\tt EVENT}: &     Activating event \\
{\tt REEP}: &     Re-entrance program pointer \\
{\tt PREV}: &     Previous suspension frame \\ \hline
\end{tabular}

\noindent
The {\tt STATE} slot indicates the current state of the frame, which can be {\it start}, {\it sleep}, {\it woken}, or {\it end}. The suspension frame enters the {\it start} state immediately after it is created and remains in it until the call is suspended for the first time or it is ended because no action rule is applicable. Normally a suspension frame transits to the {\it end} state through the {\it sleep} and {\it woken} states, but it can transit to the {\it end} state directly if its call is never suspended. The {\tt EVENT} slot stores the most recent event that activated the call. The {\tt REEP} slot stores the program pointer to continue when the call is activated. The {\tt PREV} slot stores the pointer to the previous suspension frame.

Consider, for example, the following predicate:
\begin{verbatim}
   x_in_c_y_ac(X,Y,C),var(X),var(Y),
      {dom(Y,Ey)}
       =>         
      Ex is C-Ey,        
      domain_set_false(X,Ex).
   x_in_c_y_ac(X,Y,C) => true.
\end{verbatim}
When a call {\tt x\_in\_c\_y\_ac(X,Y,C)} is executed, a suspension frame is created for it. The conditions {\tt var(X)} and {\tt var(Y)} are tested. If both are true, the frame transits from {\it start} to {\it sleep}. After an event {\tt dom(Y,Ey)} is posted, the call is activated and the state of its frame is changed from {\it sleep} to {\it woken}. All woken frames are connected into a chain of active frames by the {\tt AR} slots. When the woken call {\tt x\_in\_c\_y\_ac(X,Y,C)} is executed, the conditions {\tt var(X)} and {\tt var(Y)} are tested, and if they are true the body of the action rule is executed. If the body succeeds, the state of the suspension frame is changed from {\it woken} to {\it sleep}. If either {\tt var(X)} or {\tt var(Y)} fails, the action rule is no longer applicable and the alternative commitment rule is tried. The state of the suspension frame is changed from {\it woken} to {\it end} before the commitment rule is applied.

Events are checked at the entry and exit points of each predicate. If the queue of events is not empty, those frames that are watching the events are added into the active chain and the current predicate is interrupted. Actually, as the frame of the interrupted predicate is connected in the active chain after all the woken frames, no special action is needed to resume its execution once all the activated calls complete their execution. 

At a checkpoint for events, there may be multiple events posted that are all watched by a suspended call. If this is the case, then the frame must be copied and one copy is added into the active chain for each event. For example, if at a checkpoint two events {\tt dom(Y,Ey1)} and {\tt dom(Y,Ey2)} are on the event queue, the call {\tt x\_in\_c\_y\_ac(X,Y,C)} needs to be activated twice, one for each event. For this, the frame is copied, and the copy and the original frame are connected to the active chain, each holding one of the events in the {\tt EVENT} slot.

\begin{figure}[th]
\begin{center}
\epsfxsize=4cm \epsfbox{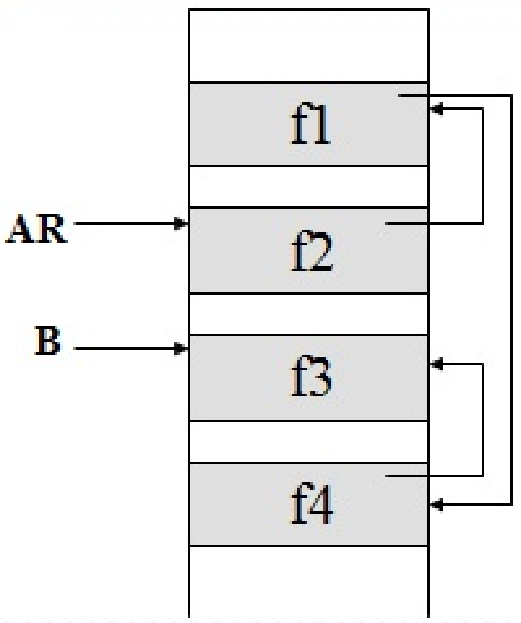}
\caption{A non-chronological active frame chain.}
\label{fig:frames}
\end{center}
\end{figure}

With suspension frames on the stack, the active chain is no longer chronological. Figure \ref{fig:frames} illustrates such a situation. The frames {\tt f1} and {\tt f2} are suspension frames, {\tt f3} is the latest choice point frame, and {\tt f4} is a determinate frame. The execution of {\tt f4} was interrupted by an event that woke up {\tt f1} and {\tt f2}. The snapshot depicts the moment immediately after the two woken frames were added into the active chain and {\tt f2} became the current active frame. 

Placing delayed calls as suspension frames on the stack makes context switching light. It is unnecessary to allocate a frame when a delayed call wakes up and deallocate it when the delayed call suspends again. Nevertheless, the non-chronologicality of the active chain on the stack requires run-time testing to determine if the current frame can be deallocated or reused. Moreover, unreachable frames on the stack need to be garbage collected \cite{Zhou00gc}. A different scheme has been proposed which stores the WAM environments for delayed calls on the heap \cite{DemoenN08}, but this scheme also complicates memory management.

\subsection{Architectural support for tabling}
B-Prolog implements a tabling mechanism, called linear tabling \cite{Zhou08tab}, which relies on iterative evaluation of top-most looping subgoals to compute fixed points. This is in contrast to the SLG mechanism implemented in XSB \cite{Sagonas98}, which relies on suspension and resumption of subgoals to compute fixed points.

A new data area, called {\it table area}, is introduced for memorizing tabled subgoals and their answers. The data structures used for the subgoal table and answer tables are orthogonal to the tabling mechanism. They can be hash tables as implemented in B-Prolog \cite{Zhou08tab}, tries as implemented in XSB \cite{Ram98}, or some other data structures. For each tabled subgoal and its variants, there is an entry in the subgoal table, which stores a pointer to the copy of the subgoal in the table, a pointer to the answer table for the subgoal, a pointer to the strongly connected component (SCC) to which the subgoal belongs, a word that indicates the state of the subgoal (e.g., whether the subgoal is complete, whether the subgoal is a looping one, and whether the answer table has been updated during the current round of evaluation). The answers in the answer table constitute a chain with a dummy answer sitting in the front. In this way, answers can be retrieved one by one through backtracking.

The frame, called a {\it tabled frame},  for a subgoal of tabled predicate contains the following two slots in addition to those slots stored in a choice point frame:   

\begin{tabular}{ll} \hline
{\tt SubgoalTable}: & Pointer to the subgoal table entry \\ 
{\tt CurrentAnswer}: & Pointer to the current answer that has been consumed \\ \hline
\end{tabular}

\noindent
The {\tt SubgoalTable} points to the subgoal table entry, and the {\tt CurrentAnswer} points to the current answer that has been consumed. The next unconsumed answer can be reached from this reference. 

When a tabled predicate is invoked by a subgoal, a tabled frame is pushed onto the stack. The subgoal table is looked up to see if a variant of the subgoal exists. If so, the {\tt SubgoalTable} slot is set to point to the entry and {\tt CurrentAnswer} is set to point to the first answer in the answer table (recall that the first answer is a dummy). If the state of the entry is complete, the subgoal only consumes existing answers one by one through backtracking. If the state of the entry is not complete, the subgoal is resolved using clauses if it appears for the first time and using existing answers if it has occurred before in the current round of evaluation. If no variant of the subgoal exists in the subgoal table, then an entry is allocated and the subgoal is resolved using clauses. 

After all clauses are tried on a tabled subgoal, a test is performed to see if the subgoal is complete. A subgoal is complete if it has never occurred in a loop, or it is a top-most looping subgoal and none of the subgoals in its SCC has obtained any new answer during the current round of evaluation. The execution of a top-most looping subgoal is iterated until it becomes complete. When a top-most looping subgoal becomes complete, all the subgoals in its SCC become complete as well.

As can be seen, the change to the architecture is minimal for supporting linear tabling. Unlike in the implementations of SLG, no effort is needed to preserve states of tabled subgoals and the garbage collector is kept untouched in linear tabling. Linear tabling is more space efficient than SLG since no stack frames are frozen for tabled subgoals. Nevertheless, linear tabling without optimization could be computationally more expensive than SLG due to the necessity of re-computation \cite{Zhou08tab}.

\section{Final Remarks}
This paper has surveyed the language features of B-Prolog and given a detailed description of TOAM Jr. with architectural support for action rules and tabling. B-Prolog has strengths and weaknesses. The competitive Prolog engine, 
the cutting-edge CLP(FD) system,
and the efficient tabling system
are clear advantages of B-Prolog. With them, B-Prolog serves well the core application domains such as constraint solving and dynamic programming. We will further strengthen B-Prolog as a tool for these applications. Future work includes parallelizing action rules for better performance in constraint solving and improving the tabling system to enhance the scalability of B-Prolog for large-scale machine-learning applications.

The shortcomings of B-Prolog are also obvious. The lack of certain functionalities such as a module system, native interfaces with database and networking libraries, and support of unicode increasingly hinders the adoption of B-Prolog in many other application domains. Additions of these new features are also part of the future work.

\section*{Acknowledgements}
Very early experiments were conducted while the author was a PhD student at Kyushu University during 1988-1991. The first working system and the versions up to 4.0 were built while the author was with Kyushu Institute of Technology during 1991-1999. Most recent improvements and enhancements have been conducted at Brooklyn College of the City University of New York. The B-Prolog system is indebted to many people in the logic programming community. I wish to express my gratitude to Taisuke Sato and Yoshitaka Kameya for their support, encouragement, and propelling. Their PRISM system has been a strong driving force for recent improvements in the tabling system and memory management. The countless feedbacks from the PRISM team greatly helped enhance the robustness of the system. Special thanks are also due to Bart Demoen for his intensive scrutiny of both the design and the implementation of B-Prolog, Yi-Dong Shen for his cooperation on linear tabling, and Paulo Moura and Ulrich Neumerkel for helping make the core part of B-Prolog more compatible with the ISO standard. Thanks also go to the anonymous referees and the editors, Maria Garc\'{i}a de la Banda and Bart Demoen, for their detailed comments and guidances on the presentation. B-Prolog-related projects have received numerous grants from various funding organizations, most recently from AIST, CISDD, PSC CUNY, and NSF.


\begin{thebibliography}{}

\bibitem[\protect\citeauthoryear{Carlsson}{Carlsson}{1987}]{Carlsson87}
{\sc Carlsson, M.} 1987.
\newblock Freeze, indexing, and other implementation issues in the {WAM}.
\newblock In {\em {Proceedings of the International Conference on Logic
  Programming (ICLP)}}. 40--58.

\bibitem[\protect\citeauthoryear{Chen and Warren}{Chen and
  Warren}{1996}]{Chen96}
{\sc Chen, W.} {\sc and} {\sc Warren, D.~S.} 1996.
\newblock Tabled evaluation with delaying for general logic programs.
\newblock {\em Journal of the ACM\/}~{\em 43,\/}~1, 20--74.

\bibitem[\protect\citeauthoryear{Debray}{Debray}{1988}]{SBProlog}
{\sc Debray, S.~K.} 1988.
\newblock {\em The {SB-Prolog} System, Version 3.0}.
\newblock SUNY Stony Brook.

\bibitem[\protect\citeauthoryear{Demoen and Nguyen}{Demoen and
  Nguyen}{2000}]{bart:wamvariations}
{\sc Demoen, B.} {\sc and} {\sc Nguyen, P.-L.} 2000.
\newblock {S}o many {WAM} variations, so little time.
\newblock In {\em Proceedings of the International Conference on Computational
  Logic (CL)}. LNAI, vol. 1861. 1240--1254.

\bibitem[\protect\citeauthoryear{Demoen and Nguyen}{Demoen and
  Nguyen}{2008a}]{DemoenN08a}
{\sc Demoen, B.} {\sc and} {\sc Nguyen, P.-L.} 2008a.
\newblock Environment reuse in the {WAM}.
\newblock In {\em {Proceedings of the International Conference on Logic
  Programming (ICLP)}}. 698--702.

\bibitem[\protect\citeauthoryear{Demoen and Nguyen}{Demoen and
  Nguyen}{2008b}]{DemoenN08}
{\sc Demoen, B.} {\sc and} {\sc Nguyen, P.-L.} 2008b.
\newblock Two {WAM} implementations of action rules.
\newblock In {\em {Proceedings of the International Conference on Logic
  Programming (ICLP)}}. 621--635.

\bibitem[\protect\citeauthoryear{Dempster, Laird, and Rubin}{Dempster
  et~al\mbox{.}}{1977}]{Dempster77}
{\sc Dempster, A.~P.}, {\sc Laird, N.~M.}, {\sc and} {\sc Rubin, D.~B.} 1977.
\newblock Maximum likelihood from incomplete data via the {EM} algorithm.
\newblock {\em Proceedings of the Royal Statistical Society\/}, 1--38.

\bibitem[\protect\citeauthoryear{Forgy}{Forgy}{1982}]{rete}
{\sc Forgy, C.~L.} 1982.
\newblock Rete: A fast algorithm for the many pattern/many object pattern match
  problem.
\newblock In {\em Artificial Intelligence}. Vol.~19. 17--37.

\bibitem[\protect\citeauthoryear{Guo and Gupta}{Guo and Gupta}{2008}]{GuoG08}
{\sc Guo, H.-F.} {\sc and} {\sc Gupta, G.} 2008.
\newblock Simplifying dynamic programming via mode-directed tabling.
\newblock {\em Softw., Pract. Exper.\/}~{\em 38,\/}~1, 75--94.

\bibitem[\protect\citeauthoryear{Hickey and Mudambi}{Hickey and
  Mudambi}{1989}]{HickeyM89}
{\sc Hickey, T.~J.} {\sc and} {\sc Mudambi, S.} 1989.
\newblock Global compilation of {Prolog}.
\newblock {\em Journal of Logic Programming\/}~{\em 7,\/}~3, 193--230.

\bibitem[\protect\citeauthoryear{Kliger and Shapiro}{Kliger and
  Shapiro}{1990}]{KligerS90}
{\sc Kliger, S.} {\sc and} {\sc Shapiro, E.~Y.} 1990.
\newblock From decision trees to decision graphs.
\newblock In {\em Proceedings of the North American Conference on Logic
  Programming (NACLP)}. 97--116.

\bibitem[\protect\citeauthoryear{Maier and Warren}{Maier and
  Warren}{1988}]{Maier88}
{\sc Maier, D.} {\sc and} {\sc Warren, D.~S.} 1988.
\newblock {\em {Computing with Logic: Logic Programming with Prolog}}.
\newblock The Benjamin/Cummings Publishing Company.

\bibitem[\protect\citeauthoryear{Meier}{Meier}{1991}]{Meier91}
{\sc Meier, M.} 1991.
\newblock Recursion versus iteration in {Prolog}.
\newblock In {\em {Proceedings of the International Conference on Logic
  Programming (ICLP)}}. 157--169.

\bibitem[\protect\citeauthoryear{Meier}{Meier}{1993}]{Meier93}
{\sc Meier, M.} 1993.
\newblock Better late than never.
\newblock In {\em ICLP-Workshop on Implementation of Logic Programming
  Systems}. 151--165.

\bibitem[\protect\citeauthoryear{Mohr and Henderson}{Mohr and
  Henderson}{1986}]{mohrhenderson:ai:1986}
{\sc Mohr, R.} {\sc and} {\sc Henderson, T.~C.} 1986.
\newblock Arc and path consistency revisited.
\newblock {\em Artificial Intelligence\/}~{\em 28}, 225--233.

\bibitem[\protect\citeauthoryear{Morales, Carro, Puebla, and
  Hermenegildo}{Morales et~al\mbox{.}}{2005}]{Morales05}
{\sc Morales, J.~F.}, {\sc Carro, M.}, {\sc Puebla, G.}, {\sc and} {\sc
  Hermenegildo, M.~V.} 2005.
\newblock A generator of efficient abstract machine implementations and its
  application to emulator minimization.
\newblock In {\em {Proceedings of the International Conference on Logic
  Programming (ICLP)}}. 21--36.

\bibitem[\protect\citeauthoryear{Moura}{Moura}{2009}]{Moura09}
{\sc Moura, P.} 2009.
\newblock From plain {Prolog} to {Logtalk} objects: Effective code
  encapsulation and reuse.
\newblock In {\em {Proceedings of the International Conference on Logic
  Programming (ICLP)}}. 23.

\bibitem[\protect\citeauthoryear{N{\"a}ss{\'e}n, Carlsson, and
  Sagonas}{N{\"a}ss{\'e}n et~al\mbox{.}}{2001}]{NassenCS01}
{\sc N{\"a}ss{\'e}n, H.}, {\sc Carlsson, M.}, {\sc and} {\sc Sagonas, K.~F.}
  2001.
\newblock Instruction merging and specialization in the {SICStus Prolog}
  virtual machine.
\newblock In {\em {Proceedings of the International Conference on Principles
  and Practice of Declarative Programming (PPDP)}}. 49--60.

\bibitem[\protect\citeauthoryear{Older and Rummell}{Older and
  Rummell}{1992}]{OlderR92}
{\sc Older, W.~J.} {\sc and} {\sc Rummell, J.~A.} 1992.
\newblock An incremental garbage collector for {WAM}-based {Prolog}.
\newblock In {\em Proceedings of the Joint International Conference and
  Symposium on Logic Programming (JICSLP)}. 369--383.

\bibitem[\protect\citeauthoryear{Ramakrishnan, Rao, Sagonas, Swift, and
  Warren}{Ramakrishnan et~al\mbox{.}}{1998}]{Ram98}
{\sc Ramakrishnan, I.}, {\sc Rao, P.}, {\sc Sagonas, K.}, {\sc Swift, T.}, {\sc
  and} {\sc Warren, D.} 1998.
\newblock Efficient access mechanisms for tabled logic programs.
\newblock {\em Journal of Logic Programming\/}~{\em 38}, 31--54.

\bibitem[\protect\citeauthoryear{Sagonas and Swift}{Sagonas and
  Swift}{1998}]{Sagonas98}
{\sc Sagonas, K.} {\sc and} {\sc Swift, T.} 1998.
\newblock An abstract machine for tabled execution of fixed-order stratified
  logic programs.
\newblock {\em ACM Transactions on Programming Languages and Systems\/}~{\em
  20,\/}~3, 586--634.

\bibitem[\protect\citeauthoryear{{Santos Costa}}{{Santos
  Costa}}{1999}]{Costa99}
{\sc {Santos Costa}, V.} 1999.
\newblock Optimizing bytecode emulation for {Prolog}.
\newblock In {\em {{Proceedings of the International Conference on Principles
  and Practice of Declarative Programming (PPDP)}}}. LNCS 1702, 261--277.

\bibitem[\protect\citeauthoryear{{Santos Costa}, Sagonas, and Lopes}{{Santos
  Costa} et~al\mbox{.}}{2007}]{CostaSL07}
{\sc {Santos Costa}, V.}, {\sc Sagonas, K.~F.}, {\sc and} {\sc Lopes, R.} 2007.
\newblock Demand-driven indexing of {Prolog} clauses.
\newblock In {\em Proceedings of the International Conference on Logic
  Programming (ICLP)}. 395--409.

\bibitem[\protect\citeauthoryear{Sato}{Sato}{2009}]{Sato09}
{\sc Sato, T.} 2009.
\newblock Generative modeling by {PRISM}.
\newblock In {\em {Proceedings of the International Conference on Logic
  Programming (ICLP)}}. 24--35.

\bibitem[\protect\citeauthoryear{Sato and Kameya}{Sato and
  Kameya}{2001}]{sato01}
{\sc Sato, T.} {\sc and} {\sc Kameya, Y.} 2001.
\newblock Parameter learning of logic programs for symbolic-statistical
  modeling.
\newblock {\em Journal of Artificial Intelligence Research\/}, 391--454.

\bibitem[\protect\citeauthoryear{Schimpf}{Schimpf}{2002}]{Schimpf02}
{\sc Schimpf, J.} 2002.
\newblock Logical loops.
\newblock In {\em {Proceedings of the International Conference on Logic
  Programming (ICLP)}}. 224--238.

\bibitem[\protect\citeauthoryear{Schrijvers, Zhou, and Demoen}{Schrijvers
  et~al\mbox{.}}{2006}]{Schrijvers06}
{\sc Schrijvers, T.}, {\sc Zhou, N.-F.}, {\sc and} {\sc Demoen, B.} 2006.
\newblock Translating constraint handling rules into action rules.
\newblock In {\em {P}roceedings of the Third Workshop on Constraint Handling
  Rules}. 141--155.

\bibitem[\protect\citeauthoryear{Tamaki and Sato}{Tamaki and
  Sato}{1986}]{Tamaki86}
{\sc Tamaki, H.} {\sc and} {\sc Sato, T.} 1986.
\newblock {OLD resolution with tabulation}.
\newblock In {\em {Proceedings of the International Conference on Logic
  Programming (ICLP)}}. 84--98.

\bibitem[\protect\citeauthoryear{Tarau and Boyer}{Tarau and
  Boyer}{1990}]{Tarau90}
{\sc Tarau, P.} {\sc and} {\sc Boyer, M.} 1990.
\newblock Elementary logic programs.
\newblock In {\em Proceedings of Programming Language Implementation and Logic
  Programming}. 159--173.

\bibitem[\protect\citeauthoryear{van Hentenryck}{van Hentenryck}{1989}]{HEN89}
{\sc van Hentenryck, P.} 1989.
\newblock {\em Constraint Satisfaction in Logic Programming}.
\newblock MIT Press.

\bibitem[\protect\citeauthoryear{{Van Roy}}{{Van Roy}}{1990}]{ROY90}
{\sc {Van Roy}, P.} 1990.
\newblock Can logic programming execute as fast as imperative programming?
\newblock {Ph.D. thesis UCB/CSD}-90-600, EECS Department, University of
  California, Berkeley.

\bibitem[\protect\citeauthoryear{{Van Roy}}{{Van Roy}}{1994}]{vanroy94}
{\sc {Van Roy}, P.} 1994.
\newblock 1983--1993: The wonder years of sequential {P}rolog implementation.
\newblock {\em Journal of Logic Programming\/}~{\em 19,20}, 385--441.

\bibitem[\protect\citeauthoryear{{Van Roy}, Demoen, and Willems}{{Van Roy}
  et~al\mbox{.}}{1987}]{RoyDW87}
{\sc {Van Roy}, P.}, {\sc Demoen, B.}, {\sc and} {\sc Willems, Y.~D.} 1987.
\newblock Improving the execution speed of compiled {Prolog} with modes, clause
  selection, and determinism.
\newblock In {\em TAPSOFT, Vol.2}. 111--125.

\bibitem[\protect\citeauthoryear{Warren}{Warren}{1977}]{WAR77}
{\sc Warren, D. H.~D.} 1977.
\newblock Implementing {Prolog}-compiling predicate logic programs.
\newblock Research report, 39-40, Dept. of Artificial Intelligence, Univ. of
  Edinburgh.

\bibitem[\protect\citeauthoryear{Warren}{Warren}{1983}]{Warren83}
{\sc Warren, D. H.~D.} 1983.
\newblock An abstract {P}rolog instruction set.
\newblock Technical note 309, SRI International.

\bibitem[\protect\citeauthoryear{Zhou}{Zhou}{1994}]{Zhou94}
{\sc Zhou, N.-F.} 1994.
\newblock On the scheme of passing arguments in stack frames for {Prolog}.
\newblock In {\em {Proceedings of the International Conference on Logic
  Programming (ICLP)}}. 159--174.

\bibitem[\protect\citeauthoryear{Zhou}{Zhou}{1996a}]{Zhou961}
{\sc Zhou, N.-F.} 1996a.
\newblock A novel implementation method of delay.
\newblock In {\em Proceedings of the Joint International Conference and
  Symposium on Logic Programming (JICSLP)}. 97--111.

\bibitem[\protect\citeauthoryear{Zhou}{Zhou}{1996b}]{Zhou96}
{\sc Zhou, N.-F.} 1996b.
\newblock Parameter passing and control stack management in {Prolog}
  implementation revisited.
\newblock {\em ACM Transactions on Programming Languages and Systems\/}~{\em
  18,\/}~6, 752--779.

\bibitem[\protect\citeauthoryear{Zhou}{Zhou}{1998}]{Zhou98}
{\sc Zhou, N.-F.} 1998.
\newblock A high-level intermediate language and the algorithms for compiling
  finite-domain constraints.
\newblock In {\em Proceedings of the Joint International Conference and
  Symposium on Logic Programming (JICSLP)}. 70--84.

\bibitem[\protect\citeauthoryear{Zhou}{Zhou}{2000}]{Zhou00gc}
{\sc Zhou, N.-F.} 2000.
\newblock Garbage collection in {B}-{P}rolog.
\newblock In {\em First Workshop on Memory Management in Logic Programming
  Implementations}. 16--25.

\bibitem[\protect\citeauthoryear{Zhou}{Zhou}{2003}]{Zhou03:cglib}
{\sc Zhou, N.-F.} 2003.
\newblock {CGLIB} --- a constraint-based graphics library.
\newblock {\em Software Practice and Experience\/}~{\em 33,\/}~13, 1199--1216.

\bibitem[\protect\citeauthoryear{Zhou}{Zhou}{2006}]{Zhou06ar}
{\sc Zhou, N.-F.} 2006.
\newblock Programming finite-domain constraint propagators in action rules.
\newblock {\em Theory and Practice of Logic Programming (TPLP)\/}~{\em 6,\/}~5,
  483--508.

\bibitem[\protect\citeauthoryear{Zhou}{Zhou}{2007}]{Zhou07toamjr}
{\sc Zhou, N.-F.} 2007.
\newblock A register-free abstract {Prolog} machine with jumbo instructions.
\newblock In {\em {Proceedings of the International Conference on Logic
  Programming (ICLP)}}. 455--457.

\bibitem[\protect\citeauthoryear{Zhou}{Zhou}{2009}]{Zhou09}
{\sc Zhou, N.-F.} 2009.
\newblock Encoding table constraints in {CLP(FD)} based on pair-wise {AC}.
\newblock In {\em {Proceedings of the International Conference on Logic
  Programming (ICLP)}}. 402--416.

\bibitem[\protect\citeauthoryear{Zhou, Kameya, and Sato}{Zhou
  et~al\mbox{.}}{2010}]{Zhou10tai}
{\sc Zhou, N.-F.}, {\sc Kameya, Y.}, {\sc and} {\sc Sato, T.} 2010.
\newblock Mode-directed tabling for dynamic programming, machine learning, and
  constraint solving.
\newblock In {\em Proceedings of the International Conference on Tools with
  Artificial Intelligence (ICTAI)}. 213--218.

\bibitem[\protect\citeauthoryear{Zhou, Sato, and Shen}{Zhou
  et~al\mbox{.}}{2008}]{Zhou08tab}
{\sc Zhou, N.-F.}, {\sc Sato, T.}, {\sc and} {\sc Shen, Y.-D.} 2008.
\newblock Linear tabling strategies and optimizations.
\newblock {\em Theory and Practice of Logic Programming (TPLP)\/}~{\em 8,\/}~1,
  81--109.

\bibitem[\protect\citeauthoryear{Zhou, Shen, and You}{Zhou
  et~al\mbox{.}}{2011}]{Zhou11LPNMR}
{\sc Zhou, N.-F.}, {\sc Shen, Y.-D.}, {\sc and} {\sc You, J.} 2011.
\newblock Compiling answer set programs into event-driven action rules.
\newblock In {\em Proceedings of the International Conference on Logic
  Programming and Nonmonotonic Reasoning (LPNMR)}. to appear.

\bibitem[\protect\citeauthoryear{Zhou, Shen, Yuan, and You}{Zhou
  et~al\mbox{.}}{2001}]{Zhou01}
{\sc Zhou, N.-F.}, {\sc Shen, Y.-D.}, {\sc Yuan, L.}, {\sc and} {\sc You, J.}
  2001.
\newblock Implementation of a linear tabling mechanism.
\newblock {\em Journal of Functional and Logic Programming\/}~{\em 2001(1)},
  1--15.

\bibitem[\protect\citeauthoryear{Zhou, Takagi, and Ushijima}{Zhou
  et~al\mbox{.}}{1990}]{ZhouTU90}
{\sc Zhou, N.-F.}, {\sc Takagi, T.}, {\sc and} {\sc Ushijima, K.} 1990.
\newblock A matching tree oriented abstract machine for {Prolog}.
\newblock In {\em {Proceedings of the International Conference on Logic
  Programming (ICLP)}}. 159--173.

\bibitem[\protect\citeauthoryear{Zhou, Wallace, and Stuckey}{Zhou
  et~al\mbox{.}}{2006}]{Zhou06tr}
{\sc Zhou, N.-F.}, {\sc Wallace, M.}, {\sc and} {\sc Stuckey, P.~J.} 2006.
\newblock The {\tt dom} event and its use in implementing constraint
  propagators.
\newblock Technical report {TR}-2006013, CUNY Compute Science.

\end{thebibliography}
\end{document}